\documentclass[aps,pra,twocolumn,showpacs,floatfix,superscriptaddress]{revtex4}
\usepackage{bm}
\usepackage{amsfonts}
\usepackage{amssymb}
\usepackage{amsmath}

\newcommand{\rd}{{\rm d}}
\newcommand{\re}{{\rm e}}
\newcommand{\ri}{{\rm i}}

\begin{document}

\title{Coordinate-space approach to vacuum polarization}

\author{Paul Indelicato}
\email[]{paul.indelicato@lkb.ens.fr}
\affiliation{\'Ecole
Normale Sup\'erieure; CNRS; Universit\'e Pierre et Marie Curie - Paris
6; 4, place Jussieu, 75252 Paris CEDEX 05, France}

\author{Peter J. Mohr}
\email[]{mohr@nist.gov}
\affiliation{National Institute of Standards and
Technology, Gaithersburg, MD 20899-8420, USA}

\author{J. Sapirstein}
\email[]{jsapirst@nd.edu}
\affiliation{University of Notre Dame, Notre Dame, IN 46556, USA} 

\date{\today}

\begin{abstract}

The vacuum-polarization correction for bound electrons or muons is
examined.  The objective is to formulate a framework for calculating the
correction from bound-state quantum electrodynamics entirely in
coordinate space, including the Uehling potential which is usually
isolated and treated separately.  Pauli-Villars regularization is
applied to the coordinate-space calculation and the most singular terms
are shown to be eliminated, leaving the physical correction after charge
renormalization.  The conventional derivation of the Uehling potential
in momentum space is reviewed and compared to the coordinate-space
derivation.

\end{abstract}
 
\maketitle
%


\section{Introduction}

In atomic hydrogen, an electron bound to a proton, the largest radiative
correction to the energy levels of S states is the self-energy
correction, which results from the electron emitting and reabsorbing a
photon.  This is in contrast to muonic hydrogen, a muon bound to a
proton, in which the vacuum polarization correction dominates.  The
reason for the difference is that for the muon, the lighter
electron-positron pair that can occur in a vacuum polarization loop
produces a large effect, while for the electron, there is no lighter
particle to produce a correspondingly large correction.  

There has been a renewed interest in the effects of vacuum polarization
due to the recent measurement of the Lamb shift in muonic hydrogen
\cite{2010108,2011079,2013011}.  As mentioned above, in this atom, the
Lamb shift is predominantly the effect of electron vacuum polarization.
However, the radius of the proton that is deduced by comparison of the
measured transition frequencies to the theoretical predictions differs
from the value obtained from spectroscopy of hydrogen and deuterium and
electron scattering experiments by 7\,$\sigma$ \cite{2012158}.  In view
of this discrepancy, a thorough review of the theory is warranted, and a
number of such investigations have been carried out, with recent reviews
given in Refs.~\cite{2013073, 2013010, 2013014, borie12, 2012044,
2011017, 2011018, 2011155}.

In this paper, we reexamine the theory of vacuum polarization.  The
objective is to provide a formulation for the calculation of the vacuum
polarization effect entirely in coordinate space for any spherically
symmetric binding field, in parallel with an earlier analysis of the
self-energy correction in coordinate space \cite{1992082}.  This
provides a framework for a direct numerical evaluation of the correction
for strong Coulomb fields and for non-Coulomb binding fields for which a
perturbation expansion is not feasible.  Vacuum polarization is
particularly problematic, because it contains the most severe
divergences of all the bound-state corrections.  

The calculation of vacuum polarization done by \citet{1956001} is also
based on a coordinate-space formulation, but it requires the explicit
analytic expression for the Green function for a point charge nucleus.
The present calculation is not based on explicit solutions, so it
provides a framework for more general potentials.  Here, the vacuum
polarization is examined in coordinate space using Pauli-Villars
regularization for an arbitrary spherically symmetric charge
distribution in the nucleus.  An alternative method of differential
regularization has also been used to treat divergences in coordinate
space \cite{1992086}.  We reprise the conventional derivation of the
Uehling potential in momentum space in order to compare it to the
coordinate-space version.

\section{Vacuum Polarization}

In bound-state QED, the second-order electron vacuum polarization
correction for an electron or muon in an external potential $V(\bm x)$
in the Feynman gauge is given, in units where $\{c\} = \{\hbar\} =
\{m_{\re}\} = 1$\cite{foot}, by (see for example \cite{1985050, 1998004})
\begin{eqnarray}
E^{(2)}_{\rm VP} &=& 4\pi \ri \alpha \int\rd(t_2-t_1)
\int {\rm d}\bm x_2 \int {\rm d}\bm x_1   \,D_{\rm F}(x_2-x_1)\,
\nonumber\\[5 pt]&&\times
 {\rm Tr}\left[ \gamma_\mu S_{\rm F}(x_2,x_2) \right]
\overline{\phi }_n (x_1) 
\gamma^\mu\phi_n(x_1)  \, .
\label{eq:vacpol}
\end{eqnarray}
In Eq.~(\ref{eq:vacpol}), $\phi_n(x)$ is a four-component wave function
given by $\phi_n(x) = \phi_n(\bm x) \, {\rm e}^{-\ri E_n t}$, where
$\phi_n(\bm x)$ is an eigenfunction of the Dirac equation
\begin{eqnarray}
\left[-\ri\bm{\alpha}\cdot\bm{\nabla}+V(\bm x)
+\beta m -E_n\right]\phi_n(\bm x) = 0  \, .
\end{eqnarray} 
Here $m$ is the bound lepton mass, the Dirac matrices are
\begin{eqnarray}
\beta &=& \left(\begin{array}{cc}I&0\\0&-I\end{array}\right)  , \qquad
\bm\alpha =\left(\begin{array}{cc}0&\bm\sigma\\
\bm\sigma&0\end{array}\right),
\end{eqnarray}
with Pauli matrices $\bm\sigma$, and the gamma matrices are $\gamma^0 =
\beta$ and $\gamma^i = \beta\alpha^i$, $i=1,2,3$. The photon propagation
function is given by
\begin{eqnarray}
D_{\rm F}(x_2-x_1)&=&-\frac{\ri}{(2\pi)^4} \int {\rm d}^4q\,\frac{e^{-\ri
q\cdot(x_2-x_1)}}{ q^2+\ri\delta}
\nonumber\\[5 pt]
&=&\frac{\ri}{8\pi^2} \int_{-\infty}^\infty {\rm d}q^0\,
\frac{\re^{-\ri q^0(t_2-t_1)} \,
\re^{-b\left|\bm x_2-\bm x_1\right|}}{\left|\bm x_2-\bm
x_1\right|} \, ,
\nonumber\\
\label{eq:pprop}
\end{eqnarray} 
with $b = -\ri\left({q^0}^2+\ri\delta\right)^{1/2}, \ {\rm Re}(b)>0$.
Integration over the time difference yields
\begin{eqnarray}
E^{(2)}_{\rm VP} &=& 
-\alpha \int \rd \bm x_2 \int \rd \bm x_1 \,
\frac{1}{\left|\bm x_2 - \bm x_1\right|}
 \nonumber\\[10 pt]&&\times
 {\rm Tr}\left[ \gamma_\mu S_{\rm F}(x_2,x_2) \right]
\overline{\phi }_n (\bm x_1) 
\gamma^\mu\phi_n(\bm x_1)  \, .
\label{eq:startex}
\end{eqnarray}
Evidently, the quantity
\begin{eqnarray}
e{\rm Tr}\left[ \gamma_0 S_{\rm F}(x_2,x_2) \right]
\end{eqnarray}
takes the role of the vacuum-polarization charge density.  The electron
propagation function can be written as
\begin{eqnarray}
S_{\rm F}(x_2,x_1)
&=&\frac{1}{2\pi \ri} \int_{-\infty}^\infty {\rm d}z\ 
G \bigl(\bm x_2, \bm x_1,z(1+\ri\delta)\bigr)
\nonumber\\[5 pt]&&\times\gamma^0 e^{-\ri z(t_2-t_1)}  \, ,
\label{eq:old37}
\end{eqnarray} 
where the Green function is the solution of the equation
\begin{eqnarray}
\left[-\ri\bm{\alpha}\cdot\bm{\nabla}_2+V(\bm x_2)
+\beta m -z\right] G(\bm x_2, \bm x_1, z) \qquad
\nonumber\\
=\delta (\bm x_2- \bm x_1) \, , 
\end{eqnarray} 
so that
\begin{eqnarray}
E^{(2)}_{\rm VP} &=& 
\frac{\ri \alpha}{2\pi} \int_{-\infty}^\infty {\rm d}z 
\int\rd \bm x_2 \int \rd \bm x_1 \, \frac{1}{|\bm x_2-\bm x_1|}
\nonumber\\[10 pt]&&\hbox to - 30 pt {} \times \big\{
{\rm Tr}\left[ G \left(\bm x_2, \bm x_2,z(1+\ri\delta)\right)\, \right]
\phi_n^\dagger(\bm x_1)\,\phi_n(\bm x_1)
\nonumber\\[10 pt]&&\hbox to - 30 pt {}
-{\rm Tr}\left[\bm\alpha \, 
G \left(\bm x_2, \bm x_2,z(1+\ri\delta)\right)\, \right]
\cdot\phi_n^\dagger(\bm x_1)\,\bm\alpha\,\phi_n(\bm x_1)
\big\} . \qquad
\label{eq:esvp}
\end{eqnarray}
For a spherically symmetric potential $V(\bm x)$, the second term
vanishes \cite{1956001}, but only after a formal cancellation of
infinite terms [see Appendix~\ref{app:vpart}].

\section{Regularization}

The expression in Eq.~(\ref{eq:esvp}) must be modified by regularization
in order to produce a valid function for either analytical or numerical
evaluation.  In particular, ${\rm Tr}\left[G\left(\bm x_2, \bm
x_2,z\right)\right]$ is meaningless, because for $\bm x_2 \approx \bm
x_1$ the trace of the Green function has a limiting form given by
\begin{eqnarray}
{\rm Tr}\left[G\left(\bm x_2, \bm x_1,z\right)\right] &=&
\frac{z}{\pi\left|\bm x_2 - \bm x_1\right|} + \dots
\end{eqnarray}
and is undefined for equal coordinates.  In addition, the integration
over $z$ is divergent.

To obtain a physical prediction from Eq.~(\ref{eq:esvp}), we carry out
Pauli-Villars regularization by replacing the Green function with the
regulated Green function given by \cite{1949014}
\begin{eqnarray}
G_{\rm R}\left(\bm x_2, \bm x_1,z\right) &=& \sum_i C_i
G_i\left(\bm x_2, \bm x_1,z\right) \, ,
\end{eqnarray}
where $G_i$ is the Green function for a Dirac particle with mass
$m_i$, the solution of
\begin{eqnarray}
\left[-\ri\bm{\alpha}\cdot\bm{\nabla}_{2}+V(\bm x_2) 
+\beta m_i - z\right] G_i(\bm x_2, \bm x_1, z) \qquad
\nonumber\\[5 pt] 
=\delta (\bm x_2 - \bm x_1)  \, ,
\end{eqnarray} 
and the $C_i$ are functions of the masses.  The leading term has
\begin{eqnarray}
C_0 &=& 1 ;\quad
m_0 = m \, .
\end{eqnarray}
The coefficients $C_i$ for $i>0$ are chosen to eliminate the divergent
terms from the level shift, and will be seen to be quotients of
polynomials in the $m_i$.  Then for $\bm x_2\ne \bm x_1$ we have the
limit
\begin{eqnarray}
\lim_{\substack{m_i\rightarrow \infty\\[1 pt]i>0}} 
G_{\rm R}\left(\bm x_2, \bm x_1,z\right) 
&=& G\left(\bm x_2, \bm x_1,z\right) \, ,
\end{eqnarray}
where $G$ is the unregulated bound-electron Green function.  The level
shift is calculated by replacing $G$ by $G_{\rm R}$ in
Eq.~(\ref{eq:esvp}) and eventually taking the limit $m_i\rightarrow
\infty$ for $i>0$ after renormalization.

\section{Expansion in $\bm{V}$}

To separate the singularities from the finite physical contribution it
is useful to employ the power series in $V$ for the Green function, an
approach that has been extensively discussed in connection with the
evaluation of QED corrections for bound states (see for example
\cite{1956001,1958003,1961010}).  The expansion is
\begin{eqnarray}
&&G\left(\bm x_2, \bm x_1,z\right) = 
F\left(\bm x_2, \bm x_1,z\right) 
\nonumber\\[10 pt]&&\quad
-\int\rd\bm w \,
F\left(\bm x_2, \bm w,z\right)V(\bm w)
F\left(\bm w, \bm x_1,z\right)
+\dots \, ,
\qquad
\end{eqnarray}
where $F$ is the free Green function, given by
\begin{eqnarray}
F(\bm x_2,\bm x_1,z) &=& 
\left[
-\ri  \bm \alpha \cdot \bm\nabla_2 + \beta m + z\right] 
\frac{\re^{-c\left|\bm x_2 - \bm x_1\right|}}
{4 \pi \left|\bm x_2 - \bm x_1\right|} \, ,
\qquad
\label{eq:free}
\end{eqnarray}
with $c = \sqrt{m^2-z^2}$, ${\rm Re}(c) > 0$. It satisfies the equation
\begin{eqnarray}
\left[-\ri\bm{\alpha}\cdot\bm{\nabla}_2
+\beta m -z\right] F(\bm x_2, \bm x_1, z)
=\delta (\bm x_2- \bm x_1) \, . \quad
\end{eqnarray} 

In this context, Furry's theorem is seen by the following consideration.
In the expansion of the electron Green function in powers of $V$, for
the term with $n$ powers of $V$, there will be altogether $n+1$
vertices and free Green functions $F$.  Thus, the expression will
consist of the trace of the sum of products of $i$ alpha matrices, $j$
beta matrices and $k$ powers of $z$, where $i+j+k=n+1$.  The trace will
vanish unless both $i$ and $j$ are even, and the integration over $z$
will vanish unless $k$ is even.  Hence only terms with $n+1$ even may be
non-zero, which is just Furry's theorem.

\subsection{Zero potential}

The leading term in the expansion is the free Green function, with the
zero-component trace given by
\begin{eqnarray}
{\rm Tr}\left[F(\bm x_2,\bm x_1,z)\right] &=& 
\frac{z\,\re^{-c\left|\bm x_2 - \bm x_1\right|}}
{\pi \left|\bm x_2 - \bm x_1\right|} \, .
\label{eq:zpt}
\end{eqnarray}
The contribution of this term to the vacuum polarization correction
formally vanishes because of Furry's theorem.  However, the term is
problematic when $\bm x_2 = \bm x_1$, which is remedied by Pauli-Villars
regularization.  In view of the analyticity of $F(\bm x_2,\bm x_1,z)$ as
a function of $z$ and the fact that its branch points are located at
$z=\pm m(1-\ri\delta)$, we can modify the contour of integration to be a
straight line along the imaginary axis plus contributions from two
quarter circles in the first and third quadrants of the complex $z$
plane.  The contribution from the two quarter circles will vanish as
their radii increase, provided the integrand of the regulated expression
falls off for large $|z|$ faster than $1/|z|$ if $\bm x_2 = \bm x_1$.
Letting $z = \ri u$, we have
\begin{eqnarray}
c_i = \sqrt{m_i^2-z^2} =
\sqrt{u^2} + \frac{m_i^2}{2\sqrt{u^2}} 
- \frac{m_i^4}{8u^2\sqrt{u^2}} + \dots \, ,
\quad
\end{eqnarray}
and the leading terms in the expansion in $1/u$ are
\begin{eqnarray}
\frac{\re^{-c_i\left|\bm x_2 - \bm x_1\right|}}
{\left|\bm x_2 - \bm x_1\right|} &=& 
\frac{\re^{-\sqrt{u^2}\,\left|\bm x_2 - \bm x_1\right|}}
{\left|\bm x_2 - \bm x_1\right|} -
\frac{m_i^2\,\re^{-\sqrt{u^2}\,\left|\bm x_2 - \bm x_1\right|}}
{2\sqrt{u^2}}
\nonumber\\[10 pt]&&+ \dots \, . \qquad
\label{eq:fex}
\end{eqnarray}
The two terms on the right-hand side of Eq.~(\ref{eq:fex}) lead to
divergent contributions, so we eliminate them with two auxiliary mass
propagators that fulfill the conditions
\begin{eqnarray}
1 + C_1 + C_2 &=& 0 \, ,
\label{eq:sum0}
\\[10 pt]
m_0^2 + C_1m_1^2 + C_2m_2^2 &=& 0 \, ,
\label{eq:sum2}
\end{eqnarray}
satisfied by
\begin{eqnarray}
C_1&=& \frac{m_0^2 - m_2^2}{m_2^2 - m_1^2} \, ,
\\[10 pt]
C_2&=& \frac{m_1^2 - m_0^2}{m_2^2 - m_1^2} \, .
\\\nonumber
\end{eqnarray}
The regularization is implemented by writing
\begin{eqnarray}
F_{\rm R}(\bm x_2,\bm x_1,z) &=&
\sum_{i=0}^2 C_i F_i(\bm x_2,\bm x_1,z) \, ,
\end{eqnarray}
where $F_i$ is given by Eq.~(\ref{eq:free}) with $m$ replaced by $m_i$.
We thus have
\begin{eqnarray}
{\rm Tr}\left[F_{\rm R}(\bm x_2,\bm x_1,z)\right] &=& 
\sum_{i=0}^2 C_i
\, \frac{\ri u\,\re^{-c_i\left|\bm x_2 - \bm x_1\right|}}
{\pi \left|\bm x_2 - \bm x_1\right|} \, ,
\end{eqnarray}
and
\begin{eqnarray}
&&{\rm Tr}\left[F_{\rm R}(\bm x_2,\bm x_2,z)\right] = 
\lim_{\left|\bm x_2 - \bm x_1\right|\rightarrow 0}
{\rm Tr}\left[F_{\rm R}(\bm x_2,\bm x_1,z)\right]
\nonumber\\[10 pt]&&\qquad\qquad= -
\frac{\ri u}{\pi}
\sum_{i=0}^2 C_i c_i
\nonumber\\[10 pt]&&\qquad\qquad= 
\frac{\ri}{8\pi u\sqrt{u^2}} \left[
\sum_{i=0}^2 C_i m_i^4 + {\cal O}\left(\frac{1}{u^2}\right)\right] .
\qquad
\end{eqnarray}
Evidently, the contribution from the quarter circles vanishes, and 
since the branches of the square root are specified to give $\sqrt{u^2}
= \left| u \right|$ for real values of $u$, the integrand is an odd
function of $u$, and we have 
\begin{eqnarray}
\int_{-\infty}^\infty \rd u \, 
{\rm Tr}\left[F_{\rm R}(\bm x_2,\bm x_2,\ri\,u)\right] &=& 0 \, .
\\\nonumber
\end{eqnarray}

\subsection{One potential}

The next term of the expansion in $V$ is
\begin{widetext}
\begin{eqnarray}
G^{(1)}\left(\bm x_2, \bm x_1,z\right) &=& 
-\int\rd\bm w \,
F\left(\bm x_2, \bm w,z\right)V(\bm w)
F\left(\bm w, \bm x_1,z\right) \, .
\label{eq:one}
\end{eqnarray}
In Appendix \ref{app:onepot} we recast ${\rm Tr}\left[G^{(1)}\left(\bm
x_2, \bm x_1,z\right)\right]$ into a form in which the singularity at
$\bm x_2 = \bm x_1$ is isolated in a relatively simple term,
\begin{eqnarray}
{\rm Tr}\left[G^{(1)}\left(\bm x_2, \bm x_1,z\right)\right] &=& 
-\frac{Z\alpha}{2\pi}
\int \rd\bm w \, 
\frac{\re^{-c\left|\bm x_2 - \bm w\right|}}
{\left|\bm x_2 - \bm w\right|}
\, \rho(\bm w) \,
\frac{\re^{-c\left|\bm w - \bm x_1\right|}}
{\left|\bm w - \bm x_1\right|}
- \frac{z^2}{2\pi^2}
\int \rd\bm w \, 
\frac{\re^{-c\left|\bm x_2 - \bm w\right|}}
{\left|\bm x_2 - \bm w\right|} \,
 V(\bm w) \,
\frac{\re^{-c\left|\bm w - \bm x_1\right|}}
{\left|\bm w - \bm x_1\right|}
\nonumber\\[10 pt]&&
- \frac{1}{2\pi}\left[V(\bm x_2) + V(\bm x_1)\right]
\frac{\re^{-c\left|\bm x_2 - \bm x_1\right|}}
{\left|\bm x_2 - \bm x_1\right|} \, ,
\qquad
\label{eq:trg1}
\end{eqnarray}
where $Z$ is the charge of the nucleus and $\rho$ is the nuclear charge
density normalized to $1$.  The limit as $\bm x_2 \rightarrow \bm x_1$
of ${\rm Tr}\left[G^{(1)}\left(\bm x_2, \bm x_1,z\right)\right]$ is
undefined due to the third term, so we introduce a counter term that
cancels the singularity pointwise and vanishes when integrated over $z$.
The counter term is
\begin{eqnarray}
G^{(1)}_{\rm A}(\bm x_2, \bm x_1, z) &=&
- \int {\rm d}\bm w \, F(\bm x_2, \bm w, z) \,
\frac{V(\bm x_2)+V(\bm x_1)}{2} \,
 F(\bm w, \bm x_1, z) \, ,
 \label{eq:ct}
\end{eqnarray}
and (see Appendix \ref{app:onepot})
\begin{eqnarray}
{\rm Tr}\left[G^{(1)}_{\rm A}(\bm x_2, \bm x_1, z)\right] &=&
- \frac{z^2}{2\pi^2}
\int \rd\bm w \, 
\frac{\re^{-c\left|\bm x_2 - \bm w\right|}}
{\left|\bm x_2 - \bm w\right|} \,
\frac{V(\bm x_2) + V(\bm x_1)}{2}\,
\frac{\re^{-c\left|\bm w - \bm x_1\right|}}
{\left|\bm w - \bm x_1\right|}
- \frac{1}{2\pi}\left[V(\bm x_2) + V(\bm x_1)\right]
\frac{\re^{-c\left|\bm x_2 - \bm x_1\right|}}
{\left|\bm x_2 - \bm x_1\right|} \, .
\qquad
\label{eq:cct}
\end{eqnarray}
The difference
\begin{eqnarray}
G^{(1)}_{\rm B}(\bm x_2, \bm x_1, z) &=&
G^{(1)}(\bm x_2, \bm x_1, z) -
G^{(1)}_{\rm A}(\bm x_2, \bm x_1, z)
\end{eqnarray}
has the trace
\begin{eqnarray}
{\rm Tr}\left[G^{(1)}_{\rm B}(\bm x_2, \bm x_1, z)\right] &=&
-\frac{Z\alpha}{2\pi}
\int \rd\bm w \, 
\frac{\re^{-c\left|\bm x_2 - \bm w\right|}}
{\left|\bm x_2 - \bm w\right|}
\, \rho(\bm w) \,
\frac{\re^{-c\left|\bm w - \bm x_1\right|}}
{\left|\bm w - \bm x_1\right|}
\nonumber\\[10 pt]&&
- \frac{z^2}{2\pi^2}
\int \rd\bm w \, 
\frac{\re^{-c\left|\bm x_2 - \bm w\right|}}
{\left|\bm x_2 - \bm w\right|}
\left[ V(\bm w) - \frac{V(\bm x_2) + V(\bm x_1)}{2}\right]
\frac{\re^{-c\left|\bm w - \bm x_1\right|}}
{\left|\bm w - \bm x_1\right|} \, ,
\qquad
\end{eqnarray}
with the limit 
\begin{eqnarray}
{\rm Tr}\left[G^{(1)}_{\rm B}(\bm x_2, \bm x_2, z)\right] &=&
-\frac{Z\alpha}{2\pi}
\int \rd\bm w \, 
\rho(\bm w) \,
\frac{\re^{-2c\left|\bm x_2 - \bm w\right|}}
{\left|\bm x_2 - \bm w\right|^2}
- \frac{z^2}{2\pi^2}
\int \rd\bm w \, 
\left[ V(\bm w) - V(\bm x_2)\right]
\frac{\re^{-2c\left|\bm x_2 - \bm w\right|}}
{\left|\bm x_2 - \bm w\right|^2}
\qquad
\end{eqnarray}
for $\bm x_1 \rightarrow \bm x_2$.

To show that the integral over the counter term vanishes, we consider
the expansion
\begin{eqnarray}
F\left(\bm x_2, \bm x_1,z+\delta z\right) &=& 
F\left(\bm x_2, \bm x_1,z\right) 
+\int\rd\bm w \,
F\left(\bm x_2, \bm w,z\right)\delta z \,
F\left(\bm w, \bm x_1,z\right)
\nonumber\\[10 pt]&&
+\int\rd\bm v \int\rd\bm w \,
F\left(\bm x_2, \bm v,z\right)\delta z \,
F\left(\bm v, \bm w,z\right)\delta z \,
F\left(\bm w, \bm x_1,z\right)
+\dots \, .
\label{eq:expansion}
\end{eqnarray}
This yields
\begin{eqnarray}
\left.\frac{\partial}{\partial \,\delta z}\,F\left(\bm x_2, \bm x_1,z+\delta
z\right)\,\right|_{\delta z = 0}
&=&
\int\rd\bm w \,
F\left(\bm x_2, \bm w,z\right)
F\left(\bm w, \bm x_1,z\right) \, ,
\end{eqnarray}
so we can write
\begin{eqnarray}
G^{(1)}_{\rm A}(\bm x_2, \bm x_1, z) &=&
-\frac{V(\bm x_2)+V(\bm x_1)}{2} \,
\frac{\partial}{\partial z}\,
 F(\bm x_2, \bm x_1, z) \, . \\ \nonumber
\end{eqnarray}
With the contour modified as in the zero-potential case, the leading
surviving term in the regulated trace is
\begin{eqnarray}
\frac{\partial}{\partial z}\,
{\rm Tr}\left[F_{\rm R}(\bm x_2,\bm x_2,z)\right] &=& 
\frac{\partial}{\partial u}\,
\frac{1}{8\pi u \sqrt{u^2}} \left[
\sum_{i=0}^2 C_i m_i^4 + {\cal O}\left(\frac{1}{u^2}\right)\right] ,
\qquad
\end{eqnarray}
so that integration gives
\begin{eqnarray}
\int_{-\infty}^\infty \rd z \, 
\frac{\partial}{\partial z}\,{\rm Tr}
\left[F_{\rm R}(\bm x_2, \bm x_1, z)\right]
&=& {\rm Tr}\left[F_{\rm R}(\bm x_2, \bm x_1, \ri
u)\right]\bigg|_{u=-\infty}^{u=\infty}
=0 \, ,
\end{eqnarray}
and the counter term vanishes.

The subtracted one-potential contribution to the Green function is
regulated to be
\begin{eqnarray}
G^{(1)}_{\rm BR}(\bm x_2, \bm x_1, z) &=& -
\int \rd\bm w \, 
\left[V(\bm w) - \frac{V(\bm x_2)+V(\bm x_1)}{2}\right] 
\sum_{i=0}^2 C_i 
F_i(\bm x_2, \bm w, z) \,
F_i(\bm w, \bm x_1, z) \, ,
\end{eqnarray}
and the trace of the corresponding equal coordinate propagation function
is
\begin{eqnarray}
&&{\rm Tr}\left[\gamma^0S_{\rm FBR}^{(1)}(x_2,x_2)\right]
=\frac{1}{2\pi \ri} \int_{-\infty}^\infty {\rm d}z\ 
{\rm Tr}\left[G^{(1)}_{\rm BR} \bigl(\bm x_2, \bm x_2,z(1+\ri\delta)\bigr)\right]
\nonumber\\[10 pt]&& \qquad\qquad =
\int_0^\infty \rd u\,
\sum_{i=0}^2 C_i 
\Bigg\{
-\frac{Z\alpha}{2\pi^2}
\int \rd\bm w \, 
\rho(\bm w) \,
\frac{\re^{-2c_i\left|\bm x_2 - \bm w\right|}}
{\left|\bm x_2 - \bm w\right|^2}
+ \frac{u^2}{2\pi^3}
\int \rd\bm w \, 
\left[ V(\bm w) - V(\bm x_2)\right]
\frac{\re^{-2c_i\left|\bm x_2 - \bm w\right|}}
{\left|\bm x_2 - \bm w\right|^2}\Bigg\} \, .
\qquad
\label{eq:sf1}
\end{eqnarray}
In this equation the second line follows from a contour rotation to the
imaginary $z$ axis and a variable change $z = \ri u$.  The contribution
from the quarter circles as $|z| \rightarrow \infty$ vanishes due to the
exponential falloff of the integrand.  For the possibly singular case of
$|\bm x_2 - \bm w| \approx 0$, we note that
\begin{eqnarray}
\sum_{i=0}^2 C_i 
\int \rd\bm w \, 
\rho(\bm w) \,
\frac{\re^{-2c_i\left|\bm x_2 - \bm w\right|}}
{\left|\bm x_2 - \bm w\right|^2}
\rightarrow
\rho(\bm x_2)
\sum_{i=0}^2 C_i
\int \rd\bm w \, 
\frac{\re^{-2c_i\left|\bm x_2 - \bm w\right|}}
{\left|\bm x_2 - \bm w\right|^2}
= \frac{3\pi}{4}
\rho(\bm x_2)
\sum_{i=0}^2 C_i \, \frac{m_i^4}{u^5}
+ {\cal O} \left(\frac{1}{u^7}\right)
\label{eq:est1}
\end{eqnarray}
and
\begin{eqnarray}
\sum_{i=0}^2 C_i \, u^2
\int \rd\bm w \, 
\left[ V(\bm w) - V(\bm x_2)\right]
\frac{\re^{-2c_i\left|\bm x_2 - \bm w\right|}}
{\left|\bm x_2 - \bm w\right|^2}
&\rightarrow&
\sum_{i=0}^2 C_i \, \frac{u^2}{6}
\left[\bm\nabla_2^2V(\bm x_2)\right]
\int \rd\bm w \, 
\re^{-2c_i\left|\bm x_2 - \bm w\right|}
\nonumber\\[10 pt]&=&
 \frac{5\pi^2 Z\alpha}{4} \, \rho(\bm x_2)
\sum_{i=0}^2 C_i \, \frac{m_i^4}{u^5}
+ {\cal O} \left(\frac{1}{u^7}\right) \, ,
\label{eq:est2}
\end{eqnarray}
which provides sufficient convergence in the integration over
$u$ for the contour rotation to be valid.

Integration by parts in Eq.~(\ref{eq:sf1}) yields
\begin{eqnarray}
&&{\rm Tr}\left[\gamma^0S_{\rm FBR}^{(1)}(x_2,x_2)\right]
\nonumber\\[10 pt]&&\qquad\qquad
= \int_0^\infty \rd u\,
\sum_{i=0}^2 C_i 
\Bigg\{
-\frac{Z\alpha u^2}{\pi^2 c_i}
\int \rd\bm w \, 
\rho(\bm w) \,
\frac{\re^{-2c_i\left|\bm x_2 - \bm w\right|}}
{\left|\bm x_2 - \bm w\right|}
+ \frac{u^4}{3\pi^3 c_i}
\int \rd\bm w \, 
\left[ V(\bm w) - V(\bm x_2)\right]
\frac{\re^{-2c_i\left|\bm x_2 - \bm w\right|}}
{\left|\bm x_2 - \bm w\right|}\Bigg\} \, ,
\qquad
\end{eqnarray}
where estimates analogous to those in Eqs.~(\ref{eq:est1}) and
(\ref{eq:est2}) show that the integration over $u$ converges and that
the surface terms at $u=\infty$ from the partial integration vanish even
for $|\bm x_2 - \bm w| \approx 0$.  For the second term we have
\begin{eqnarray}
\int \rd\bm w \, 
\left[ V(\bm w) - V(\bm x_2)\right]
\frac{\re^{-2c_i\left|\bm x_2 - \bm w\right|}}
{\left|\bm x_2 - \bm w\right|}
&=&-Z\alpha\int\rd \bm w \int \rd\bm r \, \rho(\bm r)\left(
\frac{1}{|\bm w - \bm r|} - \frac{1}{|\bm x_2 - \bm r|}\right) 
\frac{\re^{-2c_i\left|\bm x_2 - \bm w\right|}}
{\left|\bm x_2 - \bm w\right|}
\nonumber\\[10 pt]&=&
\frac{\pi Z\alpha}{c_i^2}\int \rd\bm r \, \rho(\bm r) \,
\frac{\re^{-2c_i\left|\bm x_2 - \bm r\right|}}
{\left|\bm x_2 - \bm r\right|}
\end{eqnarray}
so that
\begin{eqnarray}
{\rm Tr}\left[\gamma^0S_{\rm FBR}^{(1)}(x_2,x_2)\right]
= \frac{Z\alpha}{\pi^2}\int_0^\infty \rd u\,
\int \rd\bm r \, 
\rho(\bm r)
\sum_{i=0}^2 C_i 
\left(-\frac{u^2}{c_i} + \frac{u^4}{3c_i^3}\right)
\frac{\re^{-2c_i\left|\bm x_2 - \bm r\right|}}
{\left|\bm x_2 - \bm r\right|}\Bigg\} \, .
\label{eq:vcdcs}
\end{eqnarray}

Thus the one-potential level shift
\begin{eqnarray}
E^{(2,1)}_{\rm VP} &=& 
-\alpha
\int\rd \bm x_2 \int \rd \bm x_1 \, 
\frac{1}{\left|\bm x_2 - \bm x_1\right|} \,
{\rm Tr}\left[\gamma^0S_{\rm FBR}^{(1)}(x_2,x_2)\right]
\phi_n^\dagger(\bm x_1)\,\phi_n(\bm x_1)
\label{eq:levshift}
\end{eqnarray}
is the expectation value of a vacuum polarization potential given by
\begin{eqnarray}
V_{\rm VP}^{(2,1)}(\bm x_1) &=&
 \frac{Z\alpha^2}{\pi^2}
\int\rd \bm x_2 \, 
\frac{1}{\left|\bm x_2 - \bm x_1\right|}
\int_0^\infty \rd u 
 \int \rd \bm r \, \rho(\bm r)
\sum_{i=0}^2 C_i 
\left(\frac{u^2}{c_i} - \frac{u^4}{3c_i^3}\right)
\frac{\re^{-2c_i\left|\bm x_2 - \bm r\right|}}
{\left|\bm x_2 - \bm r\right|} \, .
\end{eqnarray}
Integration over $\bm x_2$,
\begin{eqnarray}
\int\rd \bm x_2 \, 
\frac{1}{\left|\bm x_2 - \bm x_1\right|} \,
\frac{\re^{-2c_i\left|\bm x_2 - \bm r\right|}}
{\left|\bm x_2 - \bm r\right|}
&=& \frac{\pi}{c_i^2\left|\bm x_1 - \bm r\right|}
\left(1 - \re^{-2c_i\left|\bm x_1 - \bm r\right|}\right)
\label{eq:ident}
\end{eqnarray}
yields
\begin{eqnarray}
V_{\rm VP}^{(2,1)}(\bm x_1) &=&
\frac{Z\alpha^2}{\pi}
\int_0^\infty \rd u 
 \int \rd \bm r \, \rho(\bm r)
\sum_{i=0}^2 C_i 
\left(\frac{u^2}{c_i^3} - \frac{u^4}{3c_i^5}\right)
\frac{1-\re^{-2c_i\left|\bm x_1 - \bm r\right|}}
{\left|\bm x_1 - \bm r\right|} \, .
\label{eq:vp21}
\end{eqnarray}
We have
\begin{eqnarray}
\int_0^\infty \rd u 
\sum_{i=0}^2 C_i 
\left(\frac{u^2}{c_i^3} - \frac{u^4}{3c_i^5}\right)
&=& 
-\frac{1}{3} \,
\sum_{i=0}^2 C_i 
\ln{m_i^2} \, ,
\end{eqnarray}
which produces a potential corresponding to a mass-dependent charge
proportional to the charge distribution $\rho$.  This is ultimately
eliminated by charge renormalization.  For the remaining part of
Eq.~(\ref{eq:vp21}), the contribution from each $i$ is separately finite
because of the exponential factor, and there is no contribution from the
terms with $i>0$ in the limit of large auxiliary masses.  This can be
seen from the non-relativistic estimate for the level shift from each of
these terms
\begin{eqnarray}
-\frac{Z\alpha^2}{\pi}
\,|\phi_n(0)|^2
\int\rd\bm x_1
\int_0^\infty \rd u 
 \int \rd \bm r \, \rho(\bm r)
\,C_i 
\left(\frac{u^2}{c_i^3} - \frac{u^4}{3c_i^5}\right)
\frac{\re^{-2c_i\left|\bm x_1 - \bm r\right|}}
{\left|\bm x_1 - \bm r\right|}
&=&
-Z\alpha^2
\,|\phi_n(0)|^2
\int_0^\infty \rd u 
\,C_i 
\left(\frac{u^2}{c_i^5} - \frac{u^4}{3c_i^7}\right)
\nonumber\\[10 pt]&=&
-\frac{4Z\alpha^2}{15}
\,|\phi_n(0)|^2 \,
\frac{C_i}{m_i^2}  \, .
\label{eq:nrest}
\end{eqnarray}
The surviving term of Eq.~(\ref{eq:vp21}) with $i=0$ is just the Uehling
potential $V_{\rm U}(\bm x_1)$ \cite{1935004, 1935005}.  We thus have
\begin{eqnarray}
V_{\rm VP}^{(2,1)}(\bm x_1) &=&
-\frac{Z\alpha^2}{3\pi} 
 \int \rd \bm r \, \frac{\rho(\bm r)}{\left|\bm x_1 - \bm r\right|}
\sum_{i=0}^2 C_i \, \ln{m_i^2}
+ V_{\rm U}(\bm x_1) \, ,
\label{eq:cordv}
\end{eqnarray}
where
\begin{eqnarray}
V_{\rm U}(\bm x_1) &=&
-\frac{Z\alpha^2}{\pi}
 \int \rd \bm r \, \rho(\bm r)
\int_0^\infty \rd u 
\left(\frac{u^2}{c_0^3} - \frac{u^4}{3c_0^5}\right)
\frac{\re^{-2c_0\left|\bm x_1 - \bm r\right|}}
{\left|\bm x_1 - \bm r\right|}
\nonumber\\[10 pt]&=&
-\frac{Z\alpha^2}{3\pi}
 \int \rd \bm r \, \rho(\bm r)
\int_{1}^\infty \rd t
 \, \sqrt{t^2-1}
\left( \frac{2}{t^2} + \frac{1}{t^4} \right)
\frac{\re^{-2tm_0\left|\bm x_1 - \bm r \right|}}
{\left|\bm x_1 - \bm r \right|}\, , 
\end{eqnarray}
and in the case of a point charge
\begin{eqnarray}
V_{\rm U}(\bm x_1) &=& -
\frac{Z\alpha^2}{3\pi}
\int_{1}^\infty \rd t
 \, \sqrt{t^2-1}
\left( \frac{2}{t^2} + \frac{1}{t^4} \right)
\frac{\re^{-2tm_0x_1}}
{x_1} \, ,
\label{eq:up}
\end{eqnarray}
\end{widetext}
where $x_1 = \left|\bm x_1\right|$.

To summarize, the Green function in the expression for the vacuum
polarization is not defined for equal coordinates as it appears
formally, so a counter term that removes this singularity is subtracted,
the Pauli-Villars regularization sum is made for unequal coordinates,
and the regulated expression is taken to be the limit as the coordinates
become equal.  Then, integration over the energy parameter in the Green
function is carried out, the charge is renormalized, after which the
auxiliary masses are taken to the infinite limit.  The result is just
the Uehling potential.

\begin{widetext}

\subsection{Two potential}

The next term in the expansion of the Green function in powers of $V$ is
\begin{eqnarray}
G^{(2)}\left(\bm x_2, \bm x_1,z\right) &=& 
\int\rd\bm v \int\rd\bm w \,
F\left(\bm x_2, \bm v,z\right)V(\bm v)
F\left(\bm v, \bm w,z\right)V(\bm w)
F\left(\bm w, \bm x_1,z\right) \, .
\end{eqnarray}
This gives a divergent contribution to the level shift which is expected
to vanish when regulated because of Furry's theorem.  To confirm that
the regulated contribution vanishes in our framework, we consider the
approximation given by
\begin{eqnarray}
G^{(2)}_{\rm A}\left(\bm x_2, \bm x_1,z\right) &=& 
\int\rd\bm v \int\rd\bm w \,
F\left(\bm x_2, \bm v,z\right)V(\bm x_2)
F\left(\bm v, \bm w,z\right)V(\bm x_1)
F\left(\bm w, \bm x_1,z\right) \, .
\end{eqnarray}

From Eq.~(\ref{eq:expansion}), we have
\begin{eqnarray}
\left.\frac{1}{2}\,\frac{\partial^2}{\partial \,\delta z^2}\,
F\left(\bm x_2, \bm x_1,z+\delta
z\right)\,\right|_{\delta z = 0}
&=&
\int\rd\bm v \int\rd\bm w \,
F\left(\bm x_2, \bm v,z\right)
F\left(\bm v, \bm w,z\right)
F\left(\bm w, \bm x_1,z\right) \, ,
\end{eqnarray}
so we can write
\begin{eqnarray}
G^{(2)}_{\rm A}\left(\bm x_2, \bm x_1,z\right) &=& 
\frac{1}{2}\,V(\bm x_2)V(\bm x_1)  \,
\frac{\partial^2}{\partial z^2}\,
F\left(\bm x_2, \bm x_1,z\right) \, .
\end{eqnarray}
The zero-component trace is
\begin{eqnarray}
\frac{\partial^2}{\partial z^2}\,
{\rm Tr}\left[F_i(\bm x_2, \bm x_1, z)\right]
&=&
\frac{\partial^2}{\partial z^2}\,
 \frac{z\,\re^{-c_i\left|\bm x_2 - \bm x_1\right|}}
 {\pi \left|\bm x_2 - \bm x_1\right|} 
=
- \frac{\partial^2}{\partial z^2}\,
 \frac{z\,c_i}{\pi}  
 + {\cal O}\left(\left|\bm x_2 - \bm x_1\right|\right)
\, , \qquad
\end{eqnarray}
which is finite for equal coordinates.  We thus have
\begin{eqnarray}
\frac{\partial^2}{\partial z^2}\,
{\rm Tr}\left[F_i(\bm x_2, \bm x_2, z)\right]
&=& \frac{\partial^2}{\partial u^2}\,
\frac{\ri u}{\pi}\, \sqrt{m_i^2 + u^2}
= \frac{\partial}{\partial u}\,
\frac{\ri \sqrt{u^2}}{\pi}\left(2+\frac{m_i^4}{4u^4} + \dots \right)
\qquad
\end{eqnarray}
and the regulated Green function has the factor
\begin{eqnarray}
\frac{\partial^2}{\partial z^2}\,
{\rm Tr}\left[F_{\rm R}(\bm x_2,\bm x_2,z)\right] &=& 
\frac{\partial}{\partial u}\,
\frac{\ri}{4\pi u^2 \sqrt{u^2}} \left[
\sum_{i=0}^2 C_i m_i^4 + {\cal O}\left(\frac{1}{u^2}\right)\right] ,
\qquad
\end{eqnarray}
which vanishes when integrated over $u$.  The trace of the full
regulated two-potential Green function ${\rm Tr}\left[G^{(2)}_{\rm
R}\left(\bm x_2, \bm x_1,z\right)\right]$ will have a similar analytic
behavior, and it is an odd function of $z$, so it also vanishes.

\subsection{Three potential}

The three-potential Green function is
\begin{eqnarray}
G^{(3)}\left(\bm x_2, \bm x_1,z\right) &=& 
-\int\rd\bm s\int\rd\bm v \int\rd\bm w \,
F\left(\bm x_2, \bm s,z\right)V(\bm s)
F\left(\bm s, \bm v,z\right)V(\bm v)
F\left(\bm v, \bm w,z\right)V(\bm w)
F\left(\bm w, \bm x_1,z\right) \, .
\end{eqnarray}
This expression gives the leading term in powers of $Z\alpha$ in the
all-order calculation by \citet{1956001} for a point nucleus.  For the
effect of the finite size of the nucleus on this and higher-order terms,
see \citet{1988008} and papers cited therein.  A first approximation for
this function is given by 
\begin{eqnarray}
G^{(3)}_{\rm A}\left(\bm x_2, \bm x_1,z\right) &=& 
-\int\rd\bm s\int\rd\bm v \int\rd\bm w \,
F\left(\bm x_2, \bm s,z\right)V(\bm x_2)
F\left(\bm s, \bm v,z\right)V(\bm x_2)
F\left(\bm v, \bm w,z\right)V(\bm x_2)
F\left(\bm w, \bm x_1,z\right)
\nonumber\\[10 pt] &=& 
-\frac{\left[V(\bm x_2)\right]^3}{6} \,
\frac{\partial^3}{\partial z^3}\,
F\left(\bm x_2, \bm x_1,z\right) \, .
\end{eqnarray}
The zero-component trace of the $i$th mass term is proportional to
\begin{eqnarray}
\frac{\partial^3}{\partial z^3}\,{\rm Tr}
\left[F_i(\bm x_2, \bm x_1, z)\right]
&=&
\frac{\partial^3}{\partial z^3}\,
 \frac{z\,\re^{-c_i\left|\bm x_2 - \bm x_1\right|}}
 {\pi \left|\bm x_2 - \bm x_1\right|} 
=
- \frac{\partial^3}{\partial z^3}\,
 \frac{z\,c_i}{\pi}  
 + {\cal O}\left(\left|\bm x_2 - \bm x_1\right|\right)
\, , \qquad
\end{eqnarray}
which is finite for equal coordinates, and
\begin{eqnarray}
\frac{\partial^3}{\partial z^3}\,{\rm Tr}
\left[F_i(\bm x_2, \bm x_2, z)\right]
&=& \frac{\partial^3}{\partial u^3}\,
\frac{u}{\pi}\, \sqrt{m_i^2 + u^2}
= \frac{\partial}{\partial u}\,
\frac{\sqrt{u^2}}{\pi}\left(\frac{2}{u} - 
\frac{3m_i^4}{4u^5} + \dots \right) \, .
\qquad
\end{eqnarray}
\end{widetext}

Evidently, the integral over $u$ of this expression is finite and
nonzero, because the relevant branch of the square root is positive at
$u=\pm\infty$.  This well-known property of the ``light-by-light''
Feynman diagram yields a spurious finite gauge-noninvariant part
\cite{kn}.  On the other hand, the integral of the regulated approximate
expression vanishes with no ambiguity from the quarter circles.  It is
of interest to note that the spurious contribution is not present if the
correction is calculated from an expansion of the Green function in
angular momentum eigenfunctions \cite{1988008}.  

\subsection{All-order generalization}

The higher-order terms in the potential expansion of the vacuum
polarization are finite and unambiguous, and according to Furry's
theorem, only closed loops with an even number of vertices are non-zero.
In the case considered here, there is one vertex from the interaction
with the bound electron or muon and an odd number from the expansion of
the bound Green function in powers of the external potential.

\section{Momentum-space approach}
\label{sec:msa}

Here, the conventional momentum-space derivation of the Uehling
potential for a point charge nucleus is reprised in order to provide a
comparison to the coordinate-space approach.

We consider the Fourier transform of the trace of the propagation
function in Eq.~(\ref{eq:startex}) starting from the unequal coordinate
case ${\rm Tr}\left[ \gamma^0 S_{\rm F}(x_2,x_1) \right]$.  From the
transforms
\begin{eqnarray}
F(\bm x_2, \bm x_1, z) &=& 
-\frac{1}{(2\pi)^3}
\int\rd\bm p \, 
\re^{\ri\bm p \cdot \bm x_2}
\,\frac{1}{\gamma\cdot p- m}\,\gamma^0\,
\re^{-\ri\bm p \cdot \bm x_1} \, ,
\nonumber\\
\end{eqnarray}
where $p^0 = z$, and
\begin{eqnarray}
V(\bm x) &=& -\frac{Z\alpha}{2\pi^2}\int\rd\bm p \,
\re^{\ri\bm p\cdot\bm x}\,\frac{1}{\bm p^2}
\end{eqnarray}
for the point-nucleus potential, we have from Eq.~(\ref{eq:one}) after
integration over $\bm w$
\begin{widetext}
\begin{eqnarray}
G^{(1)}\left(\bm x_2, \bm x_1,z\right) &=& 
\frac{Z\alpha}{16\pi^5} 
\int\rd\bm k \int\rd\bm p \,
\re^{\ri\bm k \cdot \bm x_2} \,
\re^{\ri\bm p \cdot (\bm x_2-\bm x_1)}
\,\frac{1}{\gamma\cdot (k + p)- m}\,\gamma^0\,
\frac{1}{\bm k^2}
\,\frac{1}{\gamma\cdot p- m}\,\gamma^0\, ,
\end{eqnarray}
where $k^0 = 0$.  The singularity for $\bm x_2 \rightarrow
\bm x_1$, as seen in Eq.~(\ref{eq:trg1}), is here manifested in the
divergence of the integral over $\bm p$ for large momenta when $\bm x_2
= \bm x_1$.  This singularity is removed by regulating with the
Pauli-Villars summation in the integrand, which gives
\begin{eqnarray}
G^{(1)}_{\rm R}\left(\bm x_2, \bm x_2,z\right) &=& 
\frac{Z\alpha}{16\pi^5} 
\int\rd\bm k \,
\frac{\re^{\ri\bm k \cdot \bm x_2}}{\bm k^2}
\int\rd\bm p \,
\sum_{i=0}^2C_i
\,\frac{1}{\gamma\cdot (k + p)- m_i}\,\gamma^0\,
\frac{1}{\gamma\cdot p- m_i}\,\gamma^0 \, .
\end{eqnarray}
We thus have
\begin{eqnarray}
{\rm Tr}\left[\gamma^0 S_{\rm FR}^{(1)}(x_2,x_2)\right] &=&
\frac{Z\alpha}{32\pi^6\ri}
\int\rd\bm k \,
\frac{\re^{\ri\bm k \cdot \bm x_2}}{\bm k^2}
\int\rd^4 p \,
\sum_{i=0}^2C_i \, {\rm Tr}\left[
\frac{1}{\gamma\cdot (k + p)- m_i}\,\gamma^0\,
\frac{1}{\gamma\cdot p- m_i}\,\gamma^0\right] \, ,
\end{eqnarray}
where it is understood that the variable $p^0$ in the integrand includes
a factor of $(1+\ri\delta)$, which is equivalent to specification of the
Feynman contour.  This expression can be written as
\begin{eqnarray}
{\rm Tr}\left[\gamma^0 S_{\rm FR}^{(1)}(x_2,x_2)\right] &=&
\frac{Z\alpha}{32\pi^6\ri}
\int\rd\bm k \,
\frac{\re^{\ri\bm k \cdot \bm x_2}}{\bm k^2} \,
I^{00}(k) \, ,
\qquad
\label{eq:cdex}
\end{eqnarray}
where
\begin{eqnarray}
I^{\mu\nu}(k) &=& \int\rd^4 p \, \sum_{i=0}^2C_i \,
{\rm Tr}\left[\gamma^\mu
\,\frac{1}{\gamma\cdot (k + p)- m_i}\,\gamma^\nu\,
\frac{1}{\gamma\cdot p- m_i}\right] \, .
\end{eqnarray}
Rotation of the contour of the variable $p^0$ to the imaginary axis and
application of standard Feynman integral evaluation methods yields (see
Appendix \ref{app:msi} for details)
\begin{eqnarray}
I^{\mu\nu}(k) &=& 
16 \pi^2 \, \ri \int_0^\infty \rd r \, r^3
\sum_{i=0}^2C_i
\int_0^1\rd y \,
\frac{\left(g^{\mu\nu}k^2-k^\mu k^\nu\right)y(1-y)}
{\left[\,r^2+\bm k^2y(1-y)+m_i^2 \right]^2}
\nonumber\\[10 pt] &&+
2 \pi^2 \, \ri \int_0^\infty \rd r \, \frac{\rd}{\rd r}\,r^4 
\sum_{i=0}^2C_i
\int_0^1\rd y \, 
\frac{g^{\mu\nu}}
{r^2+\bm k^2y(1-y)+m_i^2} \, .
\end{eqnarray}
The second term is not formally gauge invariant, but the regulated
integrand falls off sufficiently rapidly that the integral of the
derivative vanishes.  A single mass counter term regularization would
yield a non-zero result for this term.  For the first term, the change
of variable $y = (t-\sqrt{t^2-1})/2t$ and integration by parts gives
\begin{eqnarray}
\int_0^1 \rd y\,\frac{y(1-y)}
{\left[r^2 +\bm k^2y(1-y)+m_i^2 \right]^2}
&=&
\int_0^{1/2} \rd y\,\frac{2y(1-y)}
{\left[r^2 +\bm k^2y(1-y)+m_i^2 \right]^2}
\nonumber\\[10 pt]&=&
\int_1^\infty\rd t \,\frac{1}{4t^4\sqrt{t^2-1}}\,
\frac{1}{\left(r^2+\bm k^2/4t^2+m_i^2 \right)^2}
\nonumber\\[10 pt]&=& \frac{1}{6}\,
\frac{1}{\left(r^2+m_i^2 \right)^2}
-\int_1^\infty\rd t \,\frac{\sqrt{t^2-1}}{12}
\left(\frac{2}{t^4}+\frac{1}{t^6}\right)
\frac{\bm k^2}{\left(r^2+\bm k^2/4t^2+m_i^2 \right)^3} \, ,
\end{eqnarray}
and hence
\begin{eqnarray}
&&\int_0^\infty \rd r \, r^3
\sum_{i=0}^2C_i
\int_0^1 \rd y\,\frac{y(1-y)}
{\left[r^2 +\bm k^2y(1-y)+m_i^2 \right]^2}
=
-\frac{1}{12}\sum_{i=0}^2C_i \left[\ln{m_i^2}
+\int_1^\infty\rd t \, 
\, \sqrt{t^2-1}
\left(\frac{2}{t^2} + \frac{1}{t^4}\right)
\frac{\bm k^2} {\bm k^2+4t^2m_i^2} \right]
\nonumber\\[10 pt]&&\qquad\qquad\qquad\qquad\qquad=
-\frac{1}{12}\sum_{i=0}^2C_i \ln{m_i^2}
+\frac{1}{3}\int_1^\infty\rd t \, 
\, \sqrt{t^2-1}
\left(2 + \frac{1}{t^2}\right)
\sum_{i=0}^2
\frac{C_i m_i^2} {\bm k^2+4t^2m_i^2} \, .
\end{eqnarray}
Integration over $\bm k$ in Eq.~(\ref{eq:cdex}) yields
\begin{eqnarray}
{\rm Tr}\left[\gamma^0 S_{\rm FR}^{(1)}(x_2,x_2)\right] &=&
\frac{Z\alpha}{3\pi}\sum_{i=0}^2C_i\ln{m_i^2}\,\delta(\bm x_2)
-\frac{Z\alpha}{3\pi^2}\int_1^\infty\rd t \, 
\, \sqrt{t^2-1}
\left(2 + \frac{1}{t^2}\right)
\sum_{i=0}^2 C_i m_i^2 \, \frac{\re^{-2tm_ix_2}}{x_2} \, .
\label{eq:vcdms}
\end{eqnarray}

According to Eq.~(\ref{eq:levshift}), the level shift corresponds to a
potential energy given by
\begin{eqnarray}
V_{\rm VP}^{(2,1)}(\bm x_1) &=&
-\alpha\int\rd \bm x_2 \, \frac{1}{\left|\bm x_2 - \bm x_1\right|} \,
{\rm Tr}\left[\gamma^0 S_{\rm FR}^{(1)}(x_2,x_2)\right] 
\nonumber\\[10 pt]&=&
-\frac{Z\alpha^2}{3\pi}\left\{\sum_{i=0}^2C_i \, \frac{\ln{m_i^2}}{x_1}
-\int_1^\infty\rd t \, 
\, \sqrt{t^2-1}
\left(\frac{2}{t^2} + \frac{1}{t^4}\right)
\sum_{i=0}^2C_i \,
\frac{1-\re^{-2tm_ix_1}}{x_1} \right\}
\nonumber\\[10 pt]&=&
-\frac{Z\alpha^2}{3\pi}\left\{\sum_{i=0}^2C_i \, \frac{\ln{m_i^2}}{x_1}
+\int_1^\infty\rd t \, 
\, \sqrt{t^2-1}
\left(\frac{2}{t^2} + \frac{1}{t^4}\right)
\frac{\re^{-2tm_0x_1}}{x_1} \right\}  \, .
\end{eqnarray}
The term on the second line that is independent of mass vanishes from
the condition on the $C_i$ and the terms with the exponential factor and
with $i>0$ make no contribution in the large mass limit, as shown by an
estimate analogous to that in Eq.~(\ref{eq:nrest}).  The first term on
the third line is eliminated by charge renormalization and the second
term is just the Uehling potential, in agreement with Eq.~(\ref{eq:up}).

The results of the coordinate-space and momentum-space calculations
agree, as they must, since Pauli-Villars regularization makes the result
finite, but the charge renormalization terms arise in different ways.
This is seen by comparing the point charge special case of
Eq.~(\ref{eq:vcdcs})
\begin{eqnarray}
{\rm Tr}\left[\gamma^0S_{\rm FBR}^{(1)}(x_2,x_2)\right]
= \frac{Z\alpha}{\pi^2}\int_0^\infty \rd u\,
\sum_{i=0}^2 C_i 
\left(-\frac{u^2}{c_i} + \frac{u^4}{3c_i^3}\right)
\frac{\re^{-2c_ix_2}}
{x_2}
\label{eq:vcdcssc}
\end{eqnarray}
to Eq.~(\ref{eq:vcdms}).  For $x_2 \ne 0$, only the second term of
Eq.~(\ref{eq:vcdms}) is non-zero.  In Eq.~(\ref{eq:vcdcssc}), the
exponential factor provides convergence for large $u$, so that the
integrals in the sum over $i$ are separately finite.  Thus, the
mass-dependent variable change $t=c_i/m_i$ may be made for each $i$ and
the result is the same as the second term in Eq.~(\ref{eq:vcdms}).
However, if $x_2 = 0$, then the integrals over $u$ are not separately
finite and a mass-dependent variable change is not valid.  The effect of
including the point $x_2 = 0$ can be checked by integrating either
expression over $\bm x_2$.  The result of integration of
Eq.~(\ref{eq:vcdms}) is
\begin{eqnarray}
\int\rd\bm x_2 \, {\rm Tr}\left[\gamma^0S_{\rm
FR}^{(1)}(x_2,x_2)\right]
&=& 
\frac{Z\alpha}{3\pi}\sum_{i=0}^2C_i\ln{m_i^2} \, ,
\end{eqnarray}
since the integral over the second term vanishes.  Integration of
Eq.~(\ref{eq:vcdcssc}) gives
\begin{eqnarray}
\int\rd\bm x_2 \, 
{\rm Tr}\left[\gamma^0S_{\rm FBR}^{(1)}(x_2,x_2)\right]
&=& \frac{Z\alpha}{\pi}\int_0^\infty \rd u\,
\sum_{i=0}^2 C_i 
\left(-\frac{u^2}{c_i^3} + \frac{u^4}{3c_i^5}\right)
=\frac{Z\alpha}{3\pi}\sum_{i=0}^2C_i\ln{m_i^2} \, .
\label{eq:clogs}
\end{eqnarray}
\end{widetext}
These results are in agreement, although the evolution of the
logarithmic terms is quite different.

It is worth pointing out that if the (incorrect) variable change
$u\rightarrow m_iu$ were made in each term in the sum over $i$ in
Eq.~(\ref{eq:clogs}), the mass dependence would drop out and the sum
would vanish.  This illustrates the fact that the order of summation and
integration is crucial and particular care is needed in dealing with
such potentially divergent expressions.

\section{Conclusion}

Vacuum polarization for a spherically symmetric potential is examined in
coordinate space in this work.  For the contribution of first order in
the expansion in powers of the potential, a counter term is introduced
to remove the equal coordinate singularity which could be problematic
for a purely numerical calculation.  Although this singularity is
removed in principle by Pauli-Villars regularization, in practice, the
subtraction can be expected to improve the numerical convergence by
providing a pointwise removal of the singularity before numerical
integrations are carried out.  Moreover, this approach can be expected
to be useful for more general calculations.

The coordinate-space calculation is compared to the momentum-space
calculation.  With Pauli-Villars regularization, the logarithmic charge
renormalization term is seen to be the same in either case, despite the
fact that it arises in a completely different way in the two approaches.
As with the coordinate-space calculation, the momentum-space calculation
is explicitly based on the use of two Pauli-Villars auxiliary mass
subtraction terms, and it can be seen that the result is not well
defined unless both terms are included.

\appendix

\section{Vector vacuum polarization}
\label{app:vpart}

For a spherically symmetric binding potential, the vector contribution
to the vacuum polarization, in the last line of Eq.~(\ref{eq:esvp}), is
finite and vanishes when Pauli-Villars regularization is applied.  The
fact that it formally vanishes follows from the spin-angular momentum
expansion of the Green function for a spherically symmetric potential.
In particular, in this case the wave function can be written in terms
of the Dirac spin-angle functions $\chi$ as (see, for example,
\cite{1974003} and references therein)
\begin{eqnarray}
\phi_n(\bm x) &=& \left[\begin{array}{c}
f_1(x)\chi_\kappa^\mu(\bm{\hat x}) \\[10 pt]
f_2(x)\chi_{-\kappa}^\mu(\bm{\hat x})
\end{array}\right]
\end{eqnarray}
and the Green function is given by
\begin{widetext}
\begin{eqnarray}
&&G(\bm x_2,\bm x_1,z) = \sum_{\kappa\mu}
\left[\begin{array}{cc} G^{11}_\kappa(x_2,x_1,z)
\chi_\kappa^\mu (\bm{\hat x}_2)\, \chi_\kappa^{\mu\dagger}
(\bm{\hat x}_1)& -\ri\, G^{12}_\kappa(x_2,x_1,z)
\chi_\kappa^\mu (\bm{\hat x}_2)\, \chi_{-\kappa}^{\mu\dagger} (\bm{\hat x}_1)\\
\ri\,G^{21}_\kappa(x_2,x_1,z) \chi_{-\kappa}^\mu (\bm{\hat x}_2)\, 
\chi_\kappa^{\mu\dagger}(\bm{\hat x}_1)
&G^{22}_\kappa(x_2,x_1,z)\chi_{-\kappa}^\mu (\bm{\hat x}_2)\, 
\chi_{-\kappa}^{\mu\dagger} (\bm{\hat x}_1)
\end{array}\right]\, ,
\end{eqnarray} 
where
\begin{eqnarray}
G^{ij}_\kappa(x_2,x_1,z) &=& \sum_n \frac{f_i(x_2)\,f_j(x_1)}{E_n-z}
\end{eqnarray}
and
\begin{eqnarray}
\sum_\mu
\chi_\kappa^\mu (\bm{\hat x}_2)\, \chi_{\kappa}^{\mu\dagger} (\bm{\hat x}_1)
&=&
\frac{|\kappa|}{4\pi}\left[
IP_{\kappa_+}(\xi) + \frac{\ri}{\kappa} \, \bm \sigma\cdot(\bm{\hat x}_2
\times \bm{\hat x}_1)P_{\kappa_+}^{\,\prime}(\xi)
\right]
\, ;
\\[5 pt]
\sum_\mu
\chi_{-\kappa}^\mu (\bm{\hat x}_2)\, \chi_{\kappa}^{\mu\dagger} (\bm{\hat x}_1)
&=&
\frac{1}{4\pi}\,\frac{\kappa}{|\kappa|}
\left\{
\bm \sigma\cdot\bm{\hat x}_2 \,
P_{\kappa_-}^{\,\prime}(\xi)-
\bm \sigma\cdot\bm{\hat x}_1 \,
P_{\kappa_+}^{\,\prime}(\xi)
\right\} \, ,
\end{eqnarray}
with $\xi=\bm{\hat x}_2\cdot\bm{\hat x}_1$ and $\kappa_\pm =
|\kappa\pm1/2|-1/2$.  The vector trace in Eq.~(\ref{eq:esvp}) is
\begin{eqnarray}
{\rm Tr}\left[\bm \alpha \, G(\bm x_2,\bm x_1,z)\right]
&=& \frac{\ri}{2\pi}\sum_{\kappa}\frac{\kappa}{|\kappa|}\Big\{
G^{21}_\kappa(x_2,x_1,z)\left[
\bm{\hat x}_2 \,
P_{\kappa_-}^{\,\prime}(\xi)-
\bm{\hat x}_1 \,
P_{\kappa_+}^{\,\prime}(\xi)
\right] 
\nonumber\\[10 pt] &&\qquad\qquad\quad
+ G^{12}_\kappa(x_2,x_1,z)\left[
\bm{\hat x}_2 \,
P_{\kappa_+}^{\,\prime}(\xi)-
\bm{\hat x}_1 \,
P_{\kappa_-}^{\,\prime}(\xi)
\right]\Big\}
\, .
\end{eqnarray}
The equal coordinate limit is ambiguous due to the discontinuity in the
radial factor at $x_2=x_1$, but the angular factors are just
\begin{eqnarray}
\left[
\bm{\hat x}_2 \,
P_{\kappa_\pm}^{\,\prime}(\xi)-
\bm{\hat x}_1 \,
P_{\kappa_\mp}^{\,\prime}(\xi)
\right]_{\bm{\hat x}_1=\bm{\hat x}_2} &=& \pm \kappa \, \bm{\hat x}_2
\, ,
\end{eqnarray}
and integration of the level-shift expression in Eq.~(\ref{eq:esvp})
includes
\begin{eqnarray}
\int\rd\Omega_2\,\frac{\bm{\hat{x}}_2}{|\bm x_2-\bm x_1|} &=& 
\frac{4\pi}{3} \, \bm{\hat{x}}_1 \, \frac{x_<}{x_>^2} \, ,
\end{eqnarray}
so the vector term is proportional to
\begin{eqnarray}
\phi_n^\dagger(\bm x_1)\, \bm\alpha\cdot\bm{\hat x}_1 \, \phi_n(\bm x_1)
&=& 0 \, .
\end{eqnarray}

Although the vector term vanishes as shown above, the integrals giving
the level shift from the first few terms in the expansion of the Green
function in powers of the potential are not convergent.  However, these
terms are well defined when Pauli-Villars regularization is applied.
The analysis is similar to that employed in evaluating the non-vanishing
terms of the vacuum polarization and is not repeated in this case.

\section{One-potential Green function}
\label{app:onepot}

From Eqs.~(\ref{eq:free}) and (\ref{eq:one}), we have
\begin{eqnarray}
{\rm Tr}\left[G^{(1)}(\bm x_2, \bm x_1, z)\right] &=&
- \frac{1}{4\pi^2}
\left[ \bm\nabla_2 \cdot \bm\nabla_1 + m^2 + z^2\right] 
\int \rd\bm w \, 
\frac{\re^{-c\left|\bm x_2 - \bm w\right|}}
{\left|\bm x_2 - \bm w\right|}
\, V(\bm w) \,
\frac{\re^{-c\left|\bm w - \bm x_1\right|}}
{\left|\bm w - \bm x_1\right|} \, .
\qquad
\end{eqnarray}
Since $2\,\bm\nabla_2\cdot\bm\nabla_1 =
(\bm\nabla_2+\bm\nabla_1)^2-\bm\nabla_2^2-\bm\nabla_1^2$,
Eq.~(\ref{eq:trg1}) follows from
\begin{eqnarray}
\left(\bm\nabla_2 + \bm\nabla_1\right)^2
\int \rd\bm w \, 
\frac{\re^{-c\left|\bm x_2 - \bm w\right|}}
{\left|\bm x_2 - \bm w\right|}
\, V(\bm w) \,
\frac{\re^{-c\left|\bm w - \bm x_1\right|}}
{\left|\bm w - \bm x_1\right|}
&=&
\int \rd\bm w \, 
\frac{\re^{-c\left|\bm x_2 - \bm w\right|}}
{\left|\bm x_2 - \bm w\right|}
\left[\bm\nabla_w^2 \,
V(\bm w) \right]
\frac{\re^{-c\left|\bm w - \bm x_1\right|}}
{\left|\bm w - \bm x_1\right|}
\end{eqnarray}
together with
\begin{eqnarray}
\bm\nabla_w^2 \, V(\bm w) &=& 4\pi Z\alpha\,\rho(\bm w)
\end{eqnarray}
and
\begin{eqnarray}
\left(\bm\nabla_i^2-c^2\right)\frac{\re^{-c\left|\bm x_i - \bm w\right|}}
{\left|\bm x_i - \bm w\right|} &=& -4\pi\delta(\bm x_i-\bm w) \, .
\end{eqnarray}
For the counter term in Eq.~(\ref{eq:ct}), we have
\begin{eqnarray}
{\rm Tr}\left[G^{(1)}_{\rm A}(\bm x_2, \bm x_1, z)\right] &=&
- \frac{1}{4\pi^2}\left\{
\left[ \bm\nabla_2 \cdot \bm\nabla_1 + m^2 + z^2\right] 
\int \rd\bm w \, 
\frac{\re^{-c\left|\bm x_2 - \bm w\right|}}
{\left|\bm x_2 - \bm w\right|}
\frac{\re^{-c\left|\bm w - \bm x_1\right|}}
{\left|\bm w - \bm x_1\right|}\right\}
\frac{V(\bm x_2) + V(\bm x_1)}{2}
\end{eqnarray}
and Eq.~(\ref{eq:cct}) follows from
\begin{eqnarray}
\bm\nabla_2 \cdot \bm\nabla_1 
\int \rd\bm w \, 
\frac{\re^{-c\left|\bm x_2 - \bm w\right|}}
{\left|\bm x_2 - \bm w\right|}
\frac{\re^{-c\left|\bm w - \bm x_1\right|}}
{\left|\bm w - \bm x_1\right|}
&=&
\frac{2\pi}{c}\,
\bm\nabla_2 \cdot \bm\nabla_1 \,
\re^{-c|\bm x_2 - \bm x_1|}
=
\frac{2\pi}{c}
\left(\frac{2c}{|\bm x_2 - \bm x_1|} -c^2\right) 
\re^{-c|\bm x_2 - \bm x_1|}
\nonumber\\[10 pt]&=&
\frac{4\pi}{|\bm x_2 - \bm x_1|}\,
\re^{-c|\bm x_2 - \bm x_1|}
+\left(z^2-m^2\right)
\int \rd\bm w \, 
\frac{\re^{-c\left|\bm x_2 - \bm w\right|}}
{\left|\bm x_2 - \bm w\right|}
\frac{\re^{-c\left|\bm w - \bm x_1\right|}}
{\left|\bm w - \bm x_1\right|} \, .
\end{eqnarray}

\section{Momentum-space integration}
\label{app:msi}

Here we give some details of the evaluation of the function
\begin{eqnarray}
I^{\mu\nu}(k) &=&
\int\rd^4 p \, 
\sum_{i=0}^2C_i \,
{\rm Tr}\left[\gamma^\mu\,\frac{1}{\gamma\cdot (k+p)- m_i} \,
\gamma^\nu \,
\frac{1}{\gamma\cdot p- m_i} \right]
\end{eqnarray}
that appears in Sec.~\ref{sec:msa}.  Rationalization of the propagation
functions and application of the Feynman denominator formula
\begin{eqnarray}
\frac{1}{AB} &=& \int_0^1 \rd y \, \frac{1}{\left[Ay+B(1-y)\right]^2}
\end{eqnarray}
yields
\begin{eqnarray}
I^{\mu\nu}(k) &=&
\int\rd^4 p \, 
\sum_{i=0}^2C_i
\int_0^1\rd y \, \frac{
{\rm Tr}\left[\gamma^\mu\left(\gamma\cdot p+ \gamma\cdot k+ m_i\right)
\gamma^\nu \left(\gamma\cdot p+ m_i\right)\right]}
{\left[(p+ky)^2+k^2y(1-y)-m_i^2 + \ri\epsilon\,\right]^2}
\, ,
\end{eqnarray}
where $\epsilon = 2 \delta {p^0}^2$ and $\delta^2 {p^0}^2$ is
dropped.  After the translation $p\rightarrow p-ky$, the numerator is
\begin{eqnarray}
&&{\rm Tr}\left[\gamma^\mu\left(\gamma\cdot p+ 
\gamma\cdot k(1-y)+ m_i\right)
\gamma^\nu \left(\gamma\cdot p - \gamma\cdot ky + m_i\right)\right]
\nonumber\\[10 pt]&&\qquad \rightarrow 8\left(g^{\mu\nu}k^2-k^\mu
k^\nu\right)y(1-y) - 4g^{\mu\nu}\left[\,p^2+k^2y(1-y)-m_i^2\right]
+8\,p^\mu p^\nu \, ,
\end{eqnarray}
where terms odd in $p$ are not included, and the denominator is
\begin{eqnarray}
p^2+k^2y(1-y)-m_i^2+\ri\epsilon = {p^0}^2 -\bm p^2 -\bm k^2y(1-y)-m_i^2
+\ri\epsilon \, .
\end{eqnarray}
Poles of the integrand are located at
\begin{eqnarray}
p^0 = \pm\left[\bm p^2 +\bm k^2y(1-y)+m_i^2-\ri\epsilon\right]^{1/2}
\end{eqnarray}
in the second and fourth quadrants of the complex $p^0$ plane, so the
contour of the $p^0$ integration may be rotated to the imaginary axis,
and $p$ may be replaced by a Cartesian vector $q$, where $p^0 = \ri q_0$
and $p^i = q_i$ for $i=1,2,3$.  The integrals over the four-vector $p$
are thus be expressed as integrals over $r = (q_0^2 + q_1^2 + q_2^2 +
q_3^2)^{1/2}$, the magnitude of the Cartesian four-vector $q$, where
\begin{eqnarray}
\int\rd^4 p \, f(p^2)&=& 2 \pi^2 \, \ri \int_0^\infty \rd r \,r^3f(-r^2)
\, ; \\[10 pt]
\int\rd^4 p \, p^\mu p^\nu f(p^2)&=& 
-\frac{\pi^2\,\ri}{2} \int_0^\infty \rd r \, r^5 g^{\mu\nu} f(-r^2) \, .
\end{eqnarray}
We thus have
\begin{eqnarray}
I^{\mu\nu}(k) &=& 
16 \pi^2 \, \ri \int_0^\infty \rd r \, r^3
\sum_{i=0}^2C_i
\int_0^1\rd y \,
\frac{\left(g^{\mu\nu}k^2-k^\mu k^\nu\right)y(1-y)}
{\left[\,r^2+\bm k^2y(1-y)+m_i^2 \right]^2}
\nonumber\\[10 pt] &&+
2 \pi^2 \, \ri \int_0^\infty \rd r \, \frac{\rd}{\rd r}\,r^4 
\sum_{i=0}^2C_i
\int_0^1\rd y \, 
\frac{g^{\mu\nu}}
{r^2+\bm k^2y(1-y)+m_i^2} \, .
\end{eqnarray}
It is evident that the second term vanishes with Pauli-Villars
regularization from the identity
\begin{eqnarray}
\sum_{i=0}^2 \frac{C_i}{R^2+m_i^2} &=&
\frac{C_0m_1^2m_2^2 + C_1m_0^2m_2^2 + C_2m_0^2m_1^2}
{\left(R^2+m_0^2\right)\left(R^2+m_1^2\right)
\left(R^2+m_2^2\right)} \, ,
\label{eq:sumid}
\end{eqnarray}
where $R^2 = r^2+\bm k^2y(1-y)$.  Note that in the numerator on the
right-hand side of Eq.~(\ref{eq:sumid}), a possible term proportional to
$R^4$ has the vanishing coefficient $\sum_{i=0}^2C_i$ and a possible
term proportional to $R^2$ has the coefficient
\begin{eqnarray}
C_0(m_1^2+m_2^2) + C_1(m_0^2+m_2^2) + C_2(m_0^2 + m_1^2)
&=&
C_0(m_1^2+m_2^2) + C_1(m_0^2+m_2^2) + C_2(m_0^2 + m_1^2)
+\sum_{i=0}^2C_im_i^2
\nonumber\\&=&
\left(\sum_{i=0}^2C_i\right)
\left(\sum_{j=0}^2m_j^2\right)\, ,
\end{eqnarray}
which also vanishes.
\end{widetext}

\end{document}